# Bayesian Methods and Universal Darwinism

John Campbell

**Abstract.** Bayesian methods since the time of Laplace have been understood by their practitioners as closely aligned to the scientific method. Indeed a recent champion of Bayesian methods, E. T. Jaynes, titled his textbook on the subject Probability Theory: the Logic of Science. Many philosophers of science including Karl Popper and Donald Campbell have interpreted the evolution of Science as a Darwinian process consisting of a 'copy with selective retention' algorithm abstracted from Darwin's theory of Natural Selection. Arguments are presented for an isomorphism between Bayesian Methods and Darwinian processes. Universal Darwinism, as the term has been developed by Richard Dawkins, Daniel Dennett and Susan Blackmore, is the collection of scientific theories which explain the creation and evolution of their subject matter as due to the operation of Darwinian processes. These subject matters span the fields of atomic physics, chemistry, biology and the social sciences. The principle of Maximum Entropy states that systems will evolve to states of highest entropy subject to the constraints of scientific law. This principle may be inverted to provide illumination as to the nature of scientific law. Our best cosmological theories suggest the universe contained much less complexity during the period shortly after the Big Bang than it does at present. The scientific subject matter of atomic physics, chemistry, biology and the social sciences has been created since that time. An explanation is proposed for the existence of this subject matter as due to the evolution of constraints in the form of adaptations imposed on Maximum Entropy. It is argued these adaptations were discovered and instantiated through the operations of a succession of Darwinian processes.

## INTRODUCTION

Darwin understood that the algorithmic nature of natural selection suggested that its abstraction might be useful to the study of the evolution of other processes such as language. Darwinian processes, (where evolution of some subject matter is described as embodying three steps: replication, variation and differential selection) are now used as the essential explanatory mechanism in numerous fields of study in biological and social sciences such as population genetics, evolutionary epistemology, evolutionary psychology, evolutionary archaeology, memetics and evolutionary linguistics. The term Universal Darwinism has come to be applied to the collection of scientific theories utilizing Darwinian processes as their essential explanatory model (Dawkins, 1976; Blackmore, 1995; Dennett, 1995).

Prominent philosophers of science including Karl Popper and Donald Campbell have founded the active the field of Evolutionary Epistemology, the evolution of systems of knowledge (including science), in terms of a Darwinian process (Popper, 1972; Campbell, 1965).

In this light the evolution of Science might be considered in terms of copies of existing theories being made in scientists' minds, often with variations from the original. Those variations that experience preferential survival are those that are best supported by the data.

Of course 'best supported by the data' is a measure supplied by Bayesian methods. Indeed E.T. Jaynes dubbed Bayesian probability 'the logic of science'. (Jaynes, 2003).

Karl Friston and others, of the Bayesian Brain school of neuroscience, have developed a theory of mind which models aspects of mental processes as near optimal Bayesian mechanisms that update mental models from experience (Friston, 2007). Given that the brain has evolved over hundreds of millions of years, through natural selection, to gather knowledge and guide effective actions in the world, we are led to the observation that the Bayesian process of updating knowledge was discovered long before the existence of Science or even of human beings. Indeed this observation suggests that the Darwinian/Bayesian mechanism of knowledge evolution is a truly ancient and time tested strategy and, contrary to the view of post modernists, is of a different stature from other cultural accounts such as myths or religion.

In view of the homology between Darwinian and Bayesian mechanisms apparent in the evolution of Science, Section One will examine the abstract nature of Darwinian processes postulated to be operating by various theories included within universal Darwinism and will explore their Bayesian nature.

The history of the universe has involved the creation and evolution of systems possessing greater complexity over time. Much of scientific subject matter is composed of levels of complexity created and evolved since the time of the Big Bang including atomic physics, chemistry, biology and culture. It is something of a puzzle that while the history of the universe conforms to the 2nd Law of Thermodynamics many objects of scientific interest in the universe have managed to defy this overall trend and maintain their existence in states of low entropy.

The principle of Maximum Entropy tells us that systems will always evolve to states of higher entropy unless they are constrained by scientific law to do otherwise. It follows that systems will only maintain low entropy states if they are constrained to do so by scientific law. Section Two makes the argument that insight into the nature of scientific law might be gained through an understanding of the processes by which these constraints were created and evolved.

In the Discussion a recent theory from physics is examined that holds some promise of extending this paradigm to areas of scientific subject matter more fundamental than those of biology and culture.

## Darwinian Processes and Bayesian Methods

The algorithmic nature of natural selection allows its essential mechanism to be abstracted and hypothesized as a possible mechanism operating in the evolution of subject matter often other than biological. Numerous theories of this type abound in the social sciences and even in the hard sciences such as physics.

The essential abstraction from natural selection, which we are calling a Darwinian process, has been developed in the work of Richard Dawkins, Daniel Dennett and Susan Blackmore (amongst others) to consist of a three-step process:

1) Replication of system.
2) Inheritance of some characteristics that have variation amongst the offspring.
3) Differential survival of the offspring according to which variable characteristics they possess.

It has been proposed that any system adhering to this three-step algorithm, regardless of its substrate, must evolve and will evolve in the direction of an increased ability to survive (Dawkins, 1976; Dennett, 1995; Blackmore, 1995).

Offspring that inherit a slate of characteristics providing them with a substantial survival advantage will have greater reproductive success than their competitors. Characteristics that bestow greater survivability are called adaptations. Adaptations are usually discovered through processes with a random 'trial and error' component, such as genetic mutation, but the greater survivability they bestow allows them to become widespread within the population. Systems with long evolutionary histories come to accumulate many adaptations. Indeed any organism may be considered as largely an accumulation of adaptations built up over evolutionary time.

Survival of a complex system is only allowed by the 2nd Law of Thermodynamics if exchanges with the system's environment are achieved such that the entropy of the system is maintained or decreased to an extent more than made up for by the increase in environmental entropy. For instance the strategy of photosynthesis in green plants to improve their survivability (through an intricate process that transfers free energy from photons in the environment to a chemical form of energy usable by the plant) is dependent on an overall increase in entropy in the sun/earth system.

Survivability of complex systems thus depends on finding loopholes where environmental entropy increase can be used to decrease system entropy. In this sense adaptations are highly tuned to their environments and may be considered to model their environments. Henry Plotkin (1997 p.116) has argued that adaptations can be considered the knowledge of specific methods used by low entropy systems to maintain or decrease their entropy at the expense of their environments.

Bayesian probability is often defined as a measure of a state of knowledge. It is the degree of belief we should have in a proposition given the data available to us. It is a subjective measure in the sense that it is dependent on the evidence available to a given observer and may differ between observers or with a given observer over time. However it is objective in the sense that a unique probability should be calculated by all observers having the same evidence (Jaynes, 2003, pp. 44-45).

This objective nature of Bayesian probability precludes it from applying solely to the conscious minds of human observers engaged in cultural systems of knowledge such as Science. We should expect the principles of Bayesian probability to be followed by any system having internal models of external reality that are updated through experiences if those models are to be maintained in a near-optimal manner.

A central theme running through E.T. Jaynes' great text book on Bayesian probability concerns the necessary design of any robot capable of plausible reasoning. His conclusion is that such a robot can only be designed according to the principles of Bayesian probability (Jaynes, 2003). Any system of knowledge that must be updated to model changes in its environment or to model greater details of its environment is performing inference, and Bayesian probability is the unique method of performing such inference (Jaynes, 1986).

Karl Friston has argued that any adaptive system must contain internal models of its environment. He argues that the accuracy of these models in characterizing their environments is maintained by Bayesian processes that update the models through inference from data. Friston has proposed a method of understanding many aspects of cortical organization and response that he has called the free energy principle. It suggests that organisms attempt to minimize discrepancies (surprises) between their internal models and actual events in the environment (Friston, 2007). The Bayesian Brain School has produced a substantial body of research suggesting that Bayesian processes operate at the core of many unconscious mental functions to update internal models in a manner that keeps them in sync with their environments (Friston, 2007). As Friston notes, the price of experiencing surprise may often be death.

The mechanics of updating models in a Bayesian manner when new data becomes available might be understood as forming an algorithmic process:

1) Copies of the competing models brought forward along with the relevant prior data.

2) Variations in likelihoods amongst the competing models as provided by new data in accordance with Bayes' Theorem.

3) Differential survival of competing models according to their Bayesian likelihood.

It can be argued that this algorithm is isomorphic to the algorithm underlying Darwinian processes.

If we consider the range of scientific subject matter that might be included within this Darwinian/Bayesian framework of adaptive systems we should consider Science itself, the many branches of social science with 'Evolutionary' in their titles (such as Evolutionary Psychology), many aspect of behavioural and neuroscience and of course most fields of biology involving natural selection.

The primary knowledge model employed by all organisms is coded in their genetics and expressed in their phenotypes. The adaptations composing an organism form a detailed model of what is expected in their environment and how to respond to this environment in a manner facilitating survival. At the level of population genetics the composition of a population's genetics is determined by generational transformations largely mediated through natural selection (Lewontin, 1974). These transformations serve to maintain a close fit between the model inherent in the population's genome and characteristics exhibited by the population's environment.

The genetic plan retained at any given time is selected on the basis of the adaptations produced in the phenotype. Adaptations encapsulate knowledge of the environment, specifically knowledge of how to survive in the particular environment in which the organism expects to find itself (Plotkin, 1997). If we accept a definition of inference as 'conclusions drawn logically from premises' and perhaps limit the usual meanings of 'premises' and 'conclusions' to 'facts revealed by experience' and 'models derived from experience' then we might view natural selection's selective retention of adaptations as analogous to updating of knowledge through inference.

Bayesian theory tells us that there is only one mathematically sound method of updating the plausibility of models. That method is Bayes' Theorem (Jaynes 2003). Thus we should expect, to the extent that a population's genomic model maintains a close fit to its environment, that this updating is accomplished through mechanisms analogous to Bayesian methods.

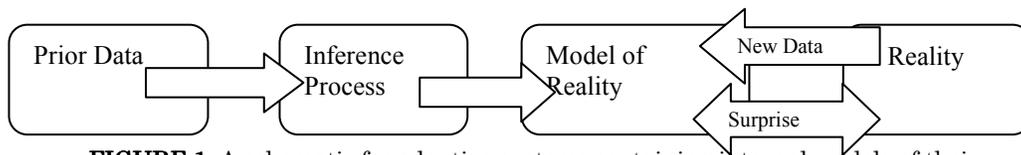

**FIGURE 1.** A schematic for adaptive systems containing internal models of their environments that are inferred, through Bayesian methods, from prior data. These models are tested by new data or experience in the environment and updated when surprised.

Since the mid-twentieth century many new fields have been introduced to the social sciences which employ Darwinian processes as explanations for the creation and evolution of their subject matter. These include evolutionary psychology, evolutionary linguistics, evolutionary archaeology, evolutionary anthropology, evolutionary sociology and evolutionary economics. These disciplines often borrow theoretical constructs directly from evolutionary biology. For instance a number of fields have adopted the biological method of cladistics, with only minor alterations. As an example Evolutionary Archeology uses cladistics in the study of artifact evolution such as arrowhead design (O'Brien, 2003). In these scenarios cultural artifacts are understood as cultural adaptations which have evolved.

Besides applying the Darwinian process to explain evolution of specific cultural entities, theories have also proposed Darwinian processes to be responsible for the overall emergence and evolution of culture itself. Two of the most developed are the Dual Inheritance theory (Henrich, 2007) and Memetics (Blackmore, 1995).

Within Universal Darwinism theories have in common the supposition that the knowledge structures of their subject matter are created and evolve through the operation of Darwinian processes. This knowledge exists in the form of adaptations that increase survivability and are passed between generations.

## Maximum Entropy and Evolution

The principle of Maximum Entropy predicts that a system will evolve to a state of highest entropy available, subject to constraints in the form of prior information applicable to the system. The constraint often represents our knowledge in the form of scientific law (Jaynes, 1985). In fact scientific law might be viewed as generalized principles that have been derived from prior information.

The second law is widely recognized as providing perhaps the most fundamental framework for the evolution of the universe. Yet when we view the actual evolution of the universe as revealed by our best scientific theories there is an apparent paradox. The scope of scientific subject matter near the time of the Big Bang was limited but since that time there have occurred successive introductions of new low-entropy subject matter including atomic physics, chemistry, biology and culture. All of these complex systems operate in accordance with the second law but they appear to exploit loopholes in order to maintain their local low entropy status.

Thermodynamic entropy in this context should be defined as the constrained maximum of $-k\,\text{Tr}(p \ln p)$ over all density matrices of the macrovariables considered (Jaynes, 1992). Thus macrovariables such as temperature impose constraints on entropy. Jaynes wrote that he could only give an answer to questions regarding the possibility of biological systems violating the second law if the variables regarding the thermodynamic state of the system were completely defined (Jaynes, 1965). For biological systems those variables would, for example, include complexities such as numerous enzymes operating in the cell serving to constrain the probability of various chemical pathways occurring and thereby constraining the entropy of the system.

The principle of Maximum Entropy tells us that the failure of complex systems that make up the bulk of scientific subject matter to evolve to states of higher entropy is due to scientific law. It is apparent that scientific law has accumulated over the course of time along with the creation of scientific subject matter. Within biology it is a near consensus that this accumulation is due to natural selection. Numerous theories within the social sciences also explain the creation and evolution of the low entropy states composing their subject matter as due to the operation of Darwinian processes.

Prior knowledge, as used by the principle of Maximum Entropy, is not only a subjective attribute of the researcher's mind. To provide useful results it must encapsulate all pertinent constraints actually occuring in nature. Indeed experimental results that do not agree with the predictions of Maximum Entropy indicate the researcher is missing pertinent prior knowledge (Jaynes, 2003, p. 371).

Thus, prior knowledge or scientific law, which forms constraints on Maximum Entropy, is an attribute of objective reality and one that has evolved during the history of the universe. An understanding of how such constraints were created and transformed may have been most thoroughly developed within biology where it forms the subject matter of evolutionary biology.

It is proposed that much of scientific law or prior knowledge may be understood as generalizations of those designs that are capable of maintaining local states of low entropy. These designs although being rare within the space of all possible designs have become widely distributed in reality due to their discovery and propagation by Darwinian processes. It is suggested that much of scientific subject matter is the accumulated adaptations discovered over evolutionary history by Darwinian processes.

Theories within Universal Darwinism describing the evolution of their subject matter in terms of information accumulated through the operation of Darwinian processes are closely aligned to the Principle of Maximum Entropy as they may explain the creation and evolution of constraints central to Maximum Entropy.

## Discussion

The unreasonable effectiveness of mathematics in the natural sciences has seemed something of a wonder to many researchers (Wigner, 1960). Why is mathematics so powerful in describing natural processes? Max Tegmark (2009) has offered a solution that may appear somewhat obvious, suggesting that physical reality is isomorphic to a mathematical structure which we are gradually discovering.

Bayesian probability may well lay claim to be the 'logic of science', given that it is the unique mathematical structure providing a valid extension of logic to areas of incomplete knowledge where degrees of plausibility must be dealt with (Jaynes, 2003, p. xx). Science is one such structure. However this claim may be extended to areas of the natural world where structures containing knowledge have evolved. Gaining knowledge requires logical inference which operates according to Bayesian principles.

Outside of the realm of biology and the social science lies much of reality that also persists in states of low entropy, notably atomic physics and chemistry. These states seem to owe their persistence as low-entropy entities directly to scientific law.

Some researchers, inspired by quantum computation, view the universe itself as a quantum computer based on the observation that the fundamental force laws of physics and their mediated interactions may be interpreted as quantum computations. The outcomes of the interactions are the results of the computations (Lloyd, 2007). Some biologists are also calling for a more information-centric focus to their subject (Brenner, 1999). We might expect that these endeavors, concerned with information and inference, will be informed by Bayesian methods and Darwinian processes.

Discovery of scientific laws might be most complete in the fields of physics and chemistry where constraints on entropy production are perhaps of a simpler design, their operation being quantum. It may seem highly unlikely that these constraints were also discovered through the action of Darwinian processes. Unfortunately there is currently a shortage of interpretations of quantum processes that are explanatory in the usual sense. Indeed Richard Feynman (1967) famously remarked that 'no one understands quantum mechanics'.

This barrier to understanding, existing for nearly a century, may have recently been breached. Wojciech Zurek, of the Los Alamos National Laboratory, and collaborators have proposed a resolution to the primary point of misunderstanding known as 'the measurement problem'. Central to this resolution is a theory they have named Quantum Darwinism. Quantum Darwinism portrays quantum interactions in terms of a Darwinian process where information is copied from the quantum system to its environment. According to this theory only a limited subset of available information can survive the transfer and forms the 'classical' reality we witness and are composed of, a reality that is selected, in a Darwinian manner, from the vast array of quantum potentialities (Zurek, 2009).

It remains to be seen if Quantum Darwinism can be successfully interpreted in a manner portraying the complexities of atomic physics and chemistry as adaptations discovered by Quantum Darwinism and following Bayesian principles.

# ACKNOWLEDGMENTS

I would like to thank Karen Drysdale, Mac Campbell, Ry Glover and Michael Skrigitil for lively and enjoyable discussions.